\documentstyle[a4,epsf]{article}
\begin{document}

\title{Quantum Computation: Particle and Wave Aspects of Algorithms}
\author{Apoorva Patel\\
{CHEP and SERC, Indian Institute of Science, Bangalore-560012}\\
{(E-mail: adpatel@cts.iisc.ernet.in)}}
\date{Special Issue in honour of Richard Feynman\\
      Resonance---Journal of Science Education,
      Vol 16, pp.821-835 (September 2011)}
\maketitle

\centerline{\bf Abstract}
\begin{center}
\parbox{15truecm}{
The driving force in the pursuit for quantum computation is the exciting
possibility that quantum algorithms can be more efficient than their
classical analogues. Research on the subject has unraveled several aspects
of how that can happen. Clever quantum algorithms have been discovered in
recent years, although not systematically, and the field remains under
active investigation. Richard Feynman was one of the pioneers who foresaw
the power of quantum computers. In this issue dedicated to him, I give
an introduction to how particle and wave aspects contribute to the power of
quantum computers. Shor's and Grover's algorithms are analysed as examples.}
\end{center}

Richard Feynman's primary area of research was quantum field theory and
particle physics. But he had many other interests beyond that (some of
them outside science), his refreshingly original approaches to analysing
different problems, and colourful ways of presenting the results to
novices as well as to the general public.
He had a life-long interest in computers, dating back to the Manhattan
project at Los Alamos while still a graduate student. There he was put
in charge of the IBM group calculating the energy release during the
implosion of the plutonium bomb, and he figured out parallel computation
techniques to speed up the work. In his later years, he became interested
in potentialities and limitations of computers, as determined by the laws
of physics, specifically quantum physics.
In the early 1980's, when I was a graduate student at Caltech, he taught
a course titled ``The Physics of Computation", together with John Hopfield
and Carver Mead. The syllabus of that course was vague, and the lecturers
covered various topics in a rather chaotic manner, often without knowing
what would come next. Still many of us attended the course for the fun of
it, especially because we knew that with Feynman as an instructor, there
would always be some surprises.
A refined version of what was taught in that course, and the exciting
ideas that developed from it, is now available as two excellent books
\cite{feyncomp1,feyncomp2}. In particular, the concept of quantum
computers formulated there has now become a thriving field of research
\cite{nielsen,archive}.

\section{Quantum Computation}

Any physical system---with some initial state, some final state, and some
interactions in between---is a candidate for an information processing device,
i.e. a computer. One only needs to construct a suitable map between the
physical properties of the system and the desired abstract mathematical
variables. The initial state becomes the input, the final state becomes
the output, and the interactions provide appropriate logic operations.
Most of the development in theoretical computer science has been in the
framework of ``particle-like" discrete digital systems. The simplest
digital language is the binary one; the digits are the bits represented
by ``off" and ``on" in electronic circuits, and by ``0" and ``1" in
mathematical notation. (Note that a language with only one letter is not
versatile enough to encode any information, but a language with two letters
can encode everything.) The outstanding advantage of a digital language is
that any value can be broken up into a sequence of digits, each one picked
from a finite set. Then $N$ possible values can be represented using only
$n=\log_2 N$ bits, which is an exponential reduction in the required
resources compared to the situation where every value is represented by
a different physical state. Mathematically this structure is known as a
``tensor product", and I will refer to similar break up of computational
algorithms as factorisation.

The growth in semiconductor technology has been so explosive---doubling
the number of transistors on a chip every 18-24 months according to Moore's
law---that many choices made in constructing the theoretical framework of
computer science (see for example, Ref.\cite{neumann}) were almost forgotten
over the years. Computer architecture became essentially synonymous with
digital electronic circuits implementing Boolean operations, pushing aside
other models of computation. It is well-known that ``wave-like" analogue
computation can also be carried out, e.g. for solving differential equations
using RLC circuits. But that has not been explored as intensively, even
though algorithms for solving differential equations on digital computers
are cumbersome due to lack of infinitesimal variables. A specific operation
may be easier to implement in the digital mode than in the analogue mode,
or vice versa, but its physical implementation is not chosen solely by
considerations of computational complexity. Any physical device cannot avoid
noise and disturbances from the environment, and hence the criteria for
hardware stability also play an important role in the choice of physical
implementation. There the discrete variables win hands down---they allow
a degree of precision, by implementation of error correction procedures for
the finite alphabet, that continuous variables cannot provide.

The situation of having to choose between digital and analogue computation
changed with the advent of quantum computation. First came the realisation
that with shrinking size of its elementary components, sooner or later,
the computer technology will inevitably encounter the dynamics of the
atomic scale \cite{feynman}. The laws that apply at the atomic scale are
those of quantum mechanics and not those of electrical circuits. The
computational framework needs reanalysis, because quantum objects display
both ``particle-like" and ``wave-like" features at the same time---the
discrete eigenstates that form the Hilbert space basis as well as the
superposition principle that allows for simultaneous existence of multiple
eigenstate components. In what way would this combination alter the axioms
of the classical information theory? How will the analysis of computational
complexity change? Many investigations in quantum information theory are
addressed to such questions. It is worthwhile to observe that this step was
precipitated by technological progress; earlier pioneers such as von Neumann,
well-versed in both quantum mechanics and computer science, had not paid any
attention to it.

The next step was automatic. In classical physics, ``particle" and ``wave"
behaviour are mutually exclusive concepts. With both ``particle-like" and
``wave-like" behaviour at their disposal, quantum algorithms can only improve
upon their classical counterparts based on only one of them.  Note that we are
classical creatures, and the inputs and the outputs of all the computational
problems we investigate are always classical (or are uniquely mapped to
classical states). At the most a quantum computer may solve a problem by
a simpler non-classical algorithmic route compared to the classical one.
We clearly understand that the concept of what is computable and what is not
does not change in going from classical to quantum computation. The scaling
rules characterising how efficiently a problem can be solved are altered,
however, and the important question is to what extent. Explorations using
several toy examples have demonstrated that the improvement provided by
a quantum solution, relative to the corresponding Boolean logic solution,
depends on the problem. The extraordinary feature is that in certain cases
the difference is large enough to challenge the conventional complexity
classification of computational problems. In what follows, we want to track
down which algorithmic advantages are due to ``particle-like" features and
which ones are due to ``wave-like" features. We use the famous algorithms
constructed by Peter Shor for factoring large integers \cite{shor}, and by
Lov Grover for database search \cite{grover}, as the test cases for our
analysis. Hopefully, the insight gained would help in developing new
quantum algorithms for other interesting computational problems.

\section{Factorisation and Superposition}

The analysis of classical computational complexity is routinely done
in the digital framework. As mentioned earlier, when the computation
can be fully factorised as a tensor product of bits, one reduces the
spatial resources needed by a factor $N/\log_2 N$. This is the maximum
gain achievable in ``particle-like" implementations; when the algorithm
does not factorise completely, the factor gained is smaller.

The characteristic property of waves useful in computation is superposition,
which means having multiple signals at the same position at the same time.
That allows the simplest type of parallel computation paradigm, i.e. SIMD
(Single Instruction Multiple Data), whereby all the superposed components
undergo identical transformations. Parallel computation generically reduces
the time complexity as the expense of the space one. For an SIMD algorithm,
the advantage depends on the number of components that can be efficiently
superposed. When a tensor product structure exists in the Hilbert space,
an exponentially large number of components can be superposed using only
polynomial resources. For instance, with $n$ qubits and $n$ rotations,
one can create a uniform superposition of $N=2^n$ components:
\begin{equation}
|0\rangle^{\otimes n} \longrightarrow
\left(\frac{|0\rangle+|1\rangle}{\sqrt{2}}\right)^{\otimes n}
= 2^{-n/2} \sum_{i=0}^{2^n-1} |i\rangle ~,
\end{equation}
where I have denoted the quantum states using Dirac's notation. Thus a
single run of a quantum algorithm can take $2^n$ superposed inputs to $2^n$
superposed outputs. The caveat is that the final measurement can extract only
one of the output components (by interference, amplification or otherwise),
while erasing all the rest. This is analogous to the situation that one
can listen to only a single radio or television programme at a time from
the superposition of a large number of available broadcast signals. The
advantage inherent in superposition is therefore useful only in those
computational problems, where many different inputs need to be processed
by the same instructions, but only one specific property of the possible
outputs is desired at the end.

The gain provided by ``wave-like" superposition is in the temporal resources.
Once again, the $N/\log_2 N$ gain, arising from superposition of $N$
components using $\log_2 N$ effort, is maximal. When the algorithm uses
a smaller number of superposed components, or when the needed superposition
cannot be created as efficiently, the factor gained is smaller.

In a general algorithm, the advantage provided by factorisation may or may
not overlap with that of superposition. One has the best of the two worlds
when the two are independent, and the gain reduces when the two overlap.
Although quantum dynamics can implement both factorisation and superposition
together, classical dynamics has to make a choice between the two, and the
digital Boolean algorithms use the former. Then the extra gain possible in
quantum algorithms, over their best Boolean counterparts, is due to the
parallelism of superposition. The superiority of a quantum algorithm thus
depends on how much superposition can be included on top of factorisation.

In the computational complexity analysis, the problems that can be solved
with all resources polynomial in the input size are considered easy, and
they form the class $P$. Alan Turing's classic work showd that a universal
computer can simulate any other computer. Subsequent analysis quantified
that the conversion cost is at most a polynomial overhead. That separated
the polynomial problems from the super-polynomial ones, irrespective of
the type of computer. Many super-polynomial problems of practical interest
belong to the class $NP$ (non-deterministic polynomial), i.e. those whose
solutions can be verified with polynomial resources. Turing's analysis did
not include superposition, and the arguments above show that superposition
can provide an exponential advantage to breach the barrier between the
classes $P$ and $NP$. That would imply that the quantum complexity analysis
of computational problems differs from the classical one. Indeed, a lot of
research effort has been directed towards discovering problems whose Boolean
solution is in the class $NP$, while the quantum solution would need only
polynomial resources, i.e. belong to the class $BQP$ (Bounded error Quantum
Polynomial time). A rigorous instance of this highly plausible conjecture
has not yet been found. But even in cases where the advantage provided by
superposition is not maximal, the quantum improvement in the scaling rules
of the algorithms can be substantial enough for practical applications.

We now look at Shor's and Grover's famous quantum algorithms in this
general framework.

\section{Shor's Algorithm \cite{shor}}

Shor's algorithm first classically reduces the problem of factoring integers
to a period finding problem, and then solves the latter using efficient
quantum implementation of Fourier transform. The discrete Quantum Fourier
Transform (QFT) can be expressed as a unitary change of basis,
\begin{equation}
\sum_x f(x) |x\rangle = \sum_y \left( {1\over\sqrt{N}}
             \sum_x e^{2\pi ixy/N} f(x) \right) |y\rangle ,
\end{equation}
again using Dirac's notation for the quantum states. Writing the integers $x$
and $y$ in binary notation, e.g. $x=x_{n-1}\cdot2^{n-1}+\ldots+x_1\cdot2+x_0$,
the non-trivial fractional part of the exponent can be decomposed as
\begin{equation}
{\rm frac}\left( {xy \over N} \right) =
y_{n-1}(.x_0)+y_{n-2}(.x_1x_0)+\ldots+y_0(.x_{n-1}\ldots x_0) .
\end{equation}
Then the unitary rotation of the QFT factorises as
\begin{equation}
|x\rangle \longrightarrow {1\over\sqrt{N}} \sum_y e^{2\pi ixy/N}|y\rangle
= {\big(|0\rangle+e^{2\pi i(.x_0)}|1\rangle\big) \over \sqrt{2}}
  {\big(|0\rangle+e^{2\pi i(.x_1x_0)}|1\rangle\big) \over \sqrt{2}}\ldots
  {\big(|0\rangle+e^{2\pi i(.x_{n-1}\ldots x_0)}|1\rangle\big) \over \sqrt{2}} ,
\end{equation}
where the sum over $y$ has been expanded in terms of the two values of each
of its $n$ bits. The factorisation has converted the sum over $N$ different
values of $y$ to a product of $n$ single qubit rotations. This is the
familiar classical trick used for implementing the Fast Fourier Transform
(FFT), albeit written in a quantum notation. Complete factorisation of the
transform provides the maximal $O(N/\log_2 N)$ gain, as expected.

The next step is to evaluate the transform for different values of $x$,
which can be implemented in parallel. Individual evaluations of $f(x)$
for each value of $x$ are not needed, however. The ``period finding" problem
requires only one result, i.e. the period, from multiple evaluations of
$f(x)$.  That is possible with maximal quantum superposition of $x$ values,
and a single run of QFT. Thus period finding using QFT gains another factor
of $N/\log_2 N$ in complexity. Since the parallelism over $x$ is totally
independent of the factorisation over $y$, both the $N/\log_2 N$ gains
can be achieved simultaneously, whereby the classical and the quantum 
algorithms differ exponentially in complexity.

To summarise, Fourier transform requires multiplication of an $N \times N$
matrix with an $N$-component vector, which is an $O(N^2)$ problem. The FFT
factorisation reduces the operations to $O(N \log_2 N)$. Furthermore, in
case of period finding, the QFT superposition cuts down the operations to
$O((\log_2 N)^2)$.

\section{Grover's Algorithm \cite{grover}}

Grover's algorithm concerns searching for a specific object in a database,
using binary oracle queries, i.e. questions of the type ``Does the selected
object have the specified property?" with only ``yes or no" answers.
This is a relativised problem, where the design of the oracle is not a
concern, and the optimal algorithm uses the minimum number or oracle queries
to locate the desired object. In absence of any structure in the database,
random pickings are as good as any other selection scheme. Then for a
database of $N$ items, each query has a success probability of $1/N$, and
on the average one requires $\langle Q\rangle=N$ queries to locate the
desired object.

The digital strategy for improving the search process is to factorise the
oracle query into smaller parts, and then sort the database in the order
of the query parts. The sorted order allows, for every query, separation
of the objects corresponding to the ``yes" answer from those corresponding
to the ``no" answer. For example, while looking for a word in a dictionary,
one first locates the first letter, then the second letter, and so on.
With discrete labels, a single binary query can uniquely identify only two
objects. A binary search tree therefore achieves maximal factorisation,
allowing the desired object to be found using $Q=\log_2 N$ queries. Note
that sorting requires significant effort, i.e. $O(N \log_2 N)$ operations
for a database of size $N$. It is the exponential change in the number of
queries for all subsequent searches that makes the laborious process of
sorting worthwhile, to be carried out once and for all.

After the oracle query factorisation, superposition can be used to
speed up the quantum search process further. But that parallelism can
only be over the possibilities addressed by each query factor, and not
over different query factors. Wave dynamics allows unique identification
of four objects using a single binary query, as illustrated in Fig.1.
So the additional gain provided by superposition is just $\log_2 4 = 2$.
Superposition is commutative, however, and that offers another advantage
that the quantum search process does not need prior sorting of the database
according to a particular order.

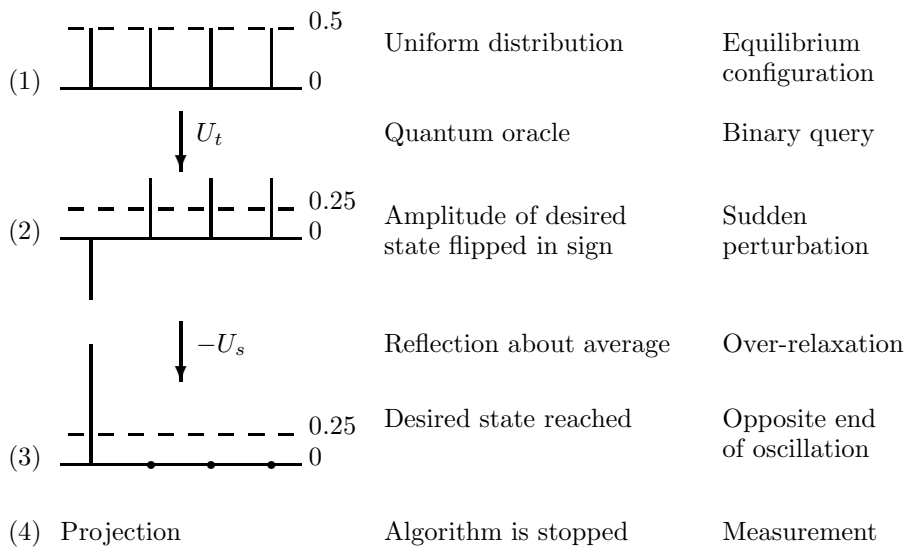
\begin{figure}[!b]
{
\setlength{\unitlength}{1mm}
\begin{picture}(120,75)
  \thicklines
  \put( 5,65){\makebox(0,0)[bl]{(1)}}
  \put(12,65){\line(1,0){32}}
\put(13,73){\line(1,0){2}} \put(17,73){\line(1,0){2}} \put(21,73){\line(1,0){2}}
\put(25,73){\line(1,0){2}} \put(29,73){\line(1,0){2}} \put(33,73){\line(1,0){2}}
\put(37,73){\line(1,0){2}} \put(41,73){\line(1,0){2}}
  \put(45,65){\makebox(0,0)[bl]{0}} \put(45,73){\makebox(0,0)[bl]{0.5}}
  \put(16,65){\line(0,1){8}} \put(24,65){\line(0,1){8}}
  \put(32,65){\line(0,1){8}} \put(40,65){\line(0,1){8}}
  \put(55,70){\makebox(0,0)[bl]{Uniform distribution}}
  \put(100,70){\makebox(0,0)[bl]{Equilibrium}}
  \put(100,66){\makebox(0,0)[bl]{configuration}}
  \put(28,62){\vector(0,-1){8}}
  \put(30,58){\makebox(0,0)[bl]{$U_t$}}
  \put(55,58){\makebox(0,0)[bl]{Quantum oracle}}
  \put(100,58){\makebox(0,0)[bl]{Binary query}}
  \put( 5,45){\makebox(0,0)[bl]{(2)}}
  \put(12,45){\line(1,0){32}}
\put(13,49){\line(1,0){2}} \put(17,49){\line(1,0){2}} \put(21,49){\line(1,0){2}}
\put(25,49){\line(1,0){2}} \put(29,49){\line(1,0){2}} \put(33,49){\line(1,0){2}}
\put(37,49){\line(1,0){2}} \put(41,49){\line(1,0){2}}
  \put(45,45){\makebox(0,0)[bl]{0}} \put(45,49){\makebox(0,0)[bl]{0.25}}
  \put(16,45){\line(0,-1){8}} \put(24,45){\line(0,1){8}}
  \put(32,45){\line(0,1){8}} \put(40,45){\line(0,1){8}}
  \put(55,47){\makebox(0,0)[bl]{Amplitude of desired}}
  \put(55,43){\makebox(0,0)[bl]{state flipped in sign}}
  \put(100,47){\makebox(0,0)[bl]{Sudden}}
  \put(100,43){\makebox(0,0)[bl]{perturbation}}
  \put(28,34){\vector(0,-1){8}}
  \put(30,30){\makebox(0,0)[bl]{$-U_s$}}
  \put(55,30){\makebox(0,0)[bl]{Reflection about average}}
  \put(100,30){\makebox(0,0)[bl]{Over-relaxation}}
  \put( 5,15){\makebox(0,0)[bl]{(3)}}
  \put(12,15){\line(1,0){32}}
\put(13,19){\line(1,0){2}} \put(17,19){\line(1,0){2}} \put(21,19){\line(1,0){2}}
\put(25,19){\line(1,0){2}} \put(29,19){\line(1,0){2}} \put(33,19){\line(1,0){2}}
\put(37,19){\line(1,0){2}} \put(41,19){\line(1,0){2}}
  \put(45,15){\makebox(0,0)[bl]{0}} \put(45,19){\makebox(0,0)[bl]{0.25}}
  \put(16,15){\line(0,1){16}}
  \put(24,15){\circle*{1}} \put(32,15){\circle*{1}} \put(40,15){\circle*{1}}
  \put(55,20){\makebox(0,0)[bl]{Desired state reached}}
  \put(100,20){\makebox(0,0)[bl]{Opposite end}}
  \put(100,16){\makebox(0,0)[bl]{of oscillation}}
  \put( 5,5){\makebox(0,0)[bl]{(4)}}
  \put(12,5){\makebox(0,0)[bl]{Projection}}
  \put(55,5){\makebox(0,0)[bl]{Algorithm is stopped}}
  \put(100,5){\makebox(0,0)[bl]{Measurement}}
\end{picture}
}
\vspace*{-1mm}
\caption{The steps of the quantum database search algorithm for the simplest
case of 4 items, when the first item is desired by the oracle. The left column
depicts the amplitudes along the 4 basis vectors, with the dashed lines showing
their average values. The middle column describes the algorithmic steps, and
the right column mentions their physical implementation in the wave language.}
\vspace*{-2mm}
\label{fig:database}
\end{figure}

Grover actually solved the quantum search problem for the situation where the
oracle query cannot be factorised. The non-factorisation restriction makes
the gain that can be achieved from superposition explicit. The individual
objects are mapped to the basis vectors of an $N$-dimensional Hilbert space.
The algorithm evolves the initial unbiased uniform superposition state
(cf. Eq.(1)) to a final state where all but the desired components vanish.
It achieves this goal by applying two reflection operations in an alternating
sequence: (i) the binary oracle query reversing the amplitude of the desired
object, and (ii) the reflection-in-the-average operation typical of a wave
oscillating about its average value. The smallest solution, i.e. $Q=1$,
is depicted in Fig.1, while more generally
\begin{equation}
(2Q+1) \sin^{-1}(1/\sqrt{N}) = \pi/2 ,
\end{equation}
and asymptotically $Q={\pi\over4}\sqrt{N}$.

Grover's algorithm does not have the SIMD structure for processing $N$
different amplitudes. Rather it needs a clever interference among the
amplitudes so that only the desired one survives at the end. The advantage
of superposition is then limited to $O(\sqrt{N})$, and is not $N/\log_2 N$.
This square-root speed-up happens to be the best one can do, as can be
inferred from the following two features of the algorithm: (i) The evolution
is restricted to the two-dimensional subspace of the Hilbert space, formed
by the initial and the final states, and hence proceeds along the shortest
path (i.e. geodesic great circle on the unitary sphere). (ii) The largest
step one can take in a given direction during any unitary evolution is
reflection, and the algorithm uses only such steps.

The superposition advantage of Grover's algorithm can be exhibited by
classical wave systems that do not have digital structure, e.g. a set
of coupled oscillators. Such systems reduce the temporal complexity (i.e.
number of queries), but the spatial complexity remains $N$ in absence
of digitisation. On the other hand, in classical digital systems with
$\log_2 N$ spatial complexity, the temporal complexity of unstructured
search remains $N$. Note that when factorised oracle queries are available,
Grover's algorithm can use them to reduce the database size in steps, e.g.
$N \rightarrow N/4 \rightarrow N/16 \rightarrow \ldots$ in case of maximal
factorisation,

To summarise, unlike the period finding problem, the search problem does
not have two factors of $N$ in its complexity that can be improved upon
independently by factorisation and superposition. Factorisation can produce
the maximal gain, but superposition cannot. The overlap between the two
limits the maximal gain to $2N/\log_2 N$, which can be looked upon as
either $N/\log_2 N \times 2$ (factorisation followed by superposition) or
$N/\sqrt{N} \times \sqrt{N}/\log_2 \sqrt{N}$ (superposition followed by
factorisation).

\section{What Else?}

The preceding two examples illustrate the advantages to be gained from
``particle-like" factorisation and ``wave-like" superposition, as well
as their interplay. The same type of analysis and inferences help in
understanding the following also.
\begin{itemize}
\itemsep=0pt
\item {\bf Simulation of quantum systems:}
An important area where quantum computation has a lot to offer is the study
of quantum models that are simplified versions of real physical systems.
Such models often help us correlate observable phenomena with appropriate
theoretical ingredients. Frequently, exact solutions are not available
even for the simplified models, and it has become commonplace to study such
models using computer simulations. Now, quantum dynamics results from the
interference among multiple quantum evolution paths---the famous double-slit
experiment being the prototype system. Classical digital simulations either
evaluate these paths one by one or approximate their sum using importance
sampling methods. It was obvious to Feynman that, for many-body quantum
systems and quantum field theories, a quantum computer can sum over these
evolutionary paths by implementing their SIMD structure as superposition
\cite{feyncomp1}. That provides an exponential gain in complexity beyond
what classical digital computers can achieve, and so would be an attractive
application for a quantum computer.
\item {\bf Spatial search:}
This is the problem where the database to be searched is spread over a number
(say $N$) of distinct locations instead of being all in one place, and there
is a locality restriction that one can proceed from any location to only its
neighbours while inspecting the objects using a binary oracle query. The
problem is interesting when the oracle query cannot be factorised, because
the locality restriction then constrains both the global operations of sorting
and superposition. The best digital classical algorithm has to inspect all
the locations one by one, which is a directed walk in some order on the
network of locations and requires $O(N)$ effort. A quantum algorithm can do
better by superposing a number of walks, but that is restricted by how fast
the walks spread. The spread obviously depends on the geometry and the
connectivity of the network, characterised for example by its dimension.

The typical method for exploring an unstructured discrete space, with the
constraint of local movements, is the random walk. In the ``particle" form,
that describes a diffusion process, which spreads according to the rule
$distance \propto \sqrt{time}$ associated with the dispersion relation
$E \propto k^2$. On the other hand, the coherent ``wave" form spreads
quadratically faster, according to the rule $distance \propto time$
associated with the dispersion relation $E \propto |k|$. Quantum spatial
search algorithms obviously use the latter, also called the quantum random
walk, and hence have the lower bound $\Omega(N^{1/d})$ for spreading over
a $d$-dimensional network. To put it differently, the best spatial search
algorithms arise in a setting that combines unitarity of quantum dynamics
with finite propagation speed of special relativity, i.e. relativistic
quantum mechanics.

Another lower bound on quantum spatial search algorithms follows from the
fact that they cannot outperform Grover's optimal algorithm which has no
restriction on movement. Combining the two bounds, the complexity of quantum
spatial search is $\Omega(N^{1/d},\sqrt{N})$. Numerical simulations verify
that these bounds can indeed be reached. It also follows that \cite{qrw}
(i) the requirement of locality weakens with increasing $d$ and Grover's
algorithm is the $d\rightarrow\infty$ limit, (ii) the clash between the two
bounds in the critical dimension $d=2$ produces logarithmic corrections to
the $\Omega(\sqrt{N})$ scaling behaviour, familiar from critical phenomena
in statistical mechanics, and (iii) the locality constraint is the strongest
in $d=1$, where quantum spatial search cannot improve upon classical spatial
search.
\item {\bf Role of entanglement:}
The classical initial and final states of our computational problems can be
digitally factorised in a specific basis. The superposition of many such
factorised states is an entangled quantum state. When the superposed
components evolve independently, e.g. in the SIMD mode as in Eq.(4),
the computational complexity gain is maximal. When the superposed components
have to interfere with each other during the course of evolution, e.g.
as in Grover's algorithm, the gain is smaller. In the design of best
quantum algorithms, therefore, what is important is the non-mixing of the
superposed components during evolution and not the amount of entanglement.
The considerations of entanglement are relegated to the ends of the
algorithms---the production of the initial superposed state, e.g. as in
Eq.(1), and the final choice of the measurement basis that extracts
appropriate results from the superposed outcomes.
\end{itemize}

To conclude, we have observed that each of ``particle-like" digital
factorisation and ``wave-like" parallelism of superposition can provide
a computational complexity gain upto $N/\log_2 N$. Traditional computer
science has extensively explored the former, but not the latter.
Quantum algorithms need to combine the two, and we do not yet have a
general framework to do that for arbitrary problems. But we believe that
the study of pure ``wave algorithms" is an excellent stepping stone to
identifying problems amenable to efficient quantum computation.

\begin{figure}[ht]
\epsfxsize=10truecm
\centerline{\epsfbox{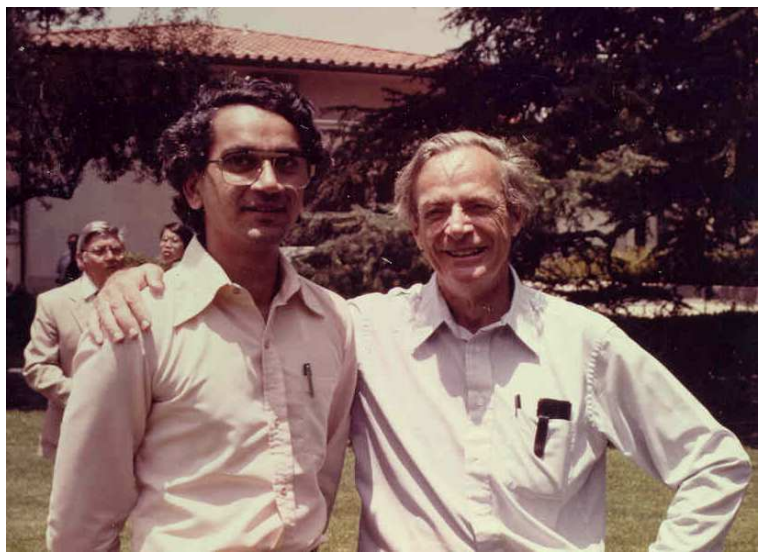}}
\caption{The author with Richard Feynman,
on his graduation day at Caltech in 1984.}
\end{figure}

\end{document}